\documentclass[aps,reprint,superscriptaddress,nofootinbib,showpacs]{revtex4-1}
\usepackage{amsmath,latexsym,amssymb,hyperref,graphicx}
\usepackage{slashed}

\hypersetup{colorlinks,citecolor= nicegreen,linkcolor= nicered}
\usepackage{color}
\definecolor{nicered}{rgb}{0.7,0.1,0.1}
\definecolor{nicegreen}{rgb}{0.1,0.5,0.1}

\usepackage{epsfig,amssymb,amsmath,latexsym,hyperref}
\usepackage{trajan}
\newcommand{\be}{\begin{equation}}
\newcommand{\ee}{\end{equation}}
\newcommand{\bea}{\begin{eqnarray}}
\newcommand{\eea}{\end{eqnarray}}
\newcommand{\no}{\noindent}

\newcommand{\de}{\partial}

\renewcommand\o{\omega}

\newcommand\m{\mu}
\newcommand\n{\nu}
\newcommand\g{\gamma}

\newcommand\ba{\begin{array}}
\newcommand\ea{\end{array}}

\newcommand\SEC[1]{\medskip\noindent{\sl\bfseries #1}}

\begin{document}
\addtolength{\belowdisplayskip}{-.3ex}       
\addtolength{\abovedisplayskip}{-.3ex}       

\title{
Finite Energy of Black Holes in Massive Gravity }

\author{Denis Comelli}
\affiliation{INFN, Sezione Ferrara, Italy}

\author{Marco Crisostomi}
\affiliation{Universit\`a dell'Aquila, L'Aquila, Italy}
\affiliation{INFN, Gruppo collegato L'Aquila, Italy}

\author{Fabrizio Nesti}
\affiliation{ICTP, Trieste, Italy}

\author{Luigi Pilo}
\affiliation{Universit\`a dell'Aquila, L'Aquila, Italy}
\affiliation{INFN, Gruppo collegato L'Aquila, Italy}

\date{\today}

\begin{abstract}
  \noindent
  In GR the static gravitational potential of a self-gravitating body
  goes as 1/r at large distances and any slower decrease leads to
  infinity energy.  We show that in a class of four-dimensional
  massive gravity theories there exists spherically symmetric
  solutions with finite total energy, featuring an asymptotic behavior
  slower than 1/r and generically of the form $r^\gamma$. This
  suggests that configurations with nonstandard asymptotics may well
  turn out to be physical.  The effect is due to an extra field
  coupled only gravitationally, which allows for modifications of the
  static potential generated by matter, while counterbalancing the
  apparently infinite energy budget.
\end{abstract}


\maketitle

\no In general relativity, the equivalence principle forbids the
localization of gravitational energy. Indeed, there exists no local
geometrical quantity that can be considered as a candidate for the
energy momentum tensor of gravity. Thus, all the numerous proposals of
local quantities, representing the gravitational energy momentum
density, are noncovariant objects ({\it pseudotensors}). Nevertheless,
it is possible to define a notion of {\it quasilocal} gravitational
energy as an integral on a closed 2-surface of a suitable covariant
density, see~\cite{Szabados:2004vb} for a review and a full list of
references. Actually, also in Newtonian gravity the total energy $E$
is determined by the static gravitational potential $\Phi$ and is
proportional to the flux of $\nabla \Phi$ on a distant sphere
surrounding the system.  Clearly, if the static potential $\Phi$
decreases at large distances slower than $1/r$ the gravitational
energy will be infinite.  

Recently, there has been a renewed interest in trying to modify
gravity at large distances trough a massive deformation of GR (for a
recent review see~\cite{Rubakov:2008nh}) extending at the nonlinear
level the seminal work of Fierz and Pauli~\cite{Fierz:1939ix}.  It was
shown that to avoid a number of phenomenological and theoretical
issues~\cite{Boulware:1972zf,vd,zak,AGS,Babichev:2010jd} Lorentz
symmetry in the gravitational sector has to be broken~\cite{rubakov,
  dubovsky, usLB,Blas}.  In this framework, an exact class of
black-hole-like solutions with striking properties were found
in~\cite{Berezhiani:2008nr}. They feature a new power-like term, in
addition to the standard newtonian one, i.e.\ $g_{tt}=1-2M/r+2S\,
r^\gamma$, where $S$ is a new integration constant.  While the energy
of standard Schwarzschild solution coincides with the mass parameter
$M$, when $\g>-1$ one would expect that the total energy of the
solution is infinite.  More in general, it is interesting to
investigate whether a solution of the Einstein equations that falls
off at infinity slower than $1/r$ can still have finite energy.

To address this question, it is important to realize that to turn the
massless spin two graviton into a massive one requires the addition of
new degrees of freedom (DoF). Following an effective field theory
approach, one can consider GR plus a minimal set of new fields
(usually called St\"uckelberg fields, borrowing the name from particle
physics), that provide the needed DoF.  The model building rules are
based on symmetries, without bothering about the dynamical origin of
the new DoF.  On the other hand, if a full theory is provided, the
effective theory can be viewed as the limit in which all the heavy
modes are decoupled, leaving only the light ones.  This point of view
is well known in particle physics; for instance, an effective version
of the standard model in the broken phase can be safely obtained by
sending to infinity the mass of the Higgs particle, retaining only the
would-be Goldstone bosons and the gauge fields.  In the case of
gravity however, playing naively the same game is dangerous because
the heavy modes also gravitate, and removing them can spoil the energy
balance of the full system.

In this letter we address the problem of the total energy for a new
class of black-hole solutions found in massive gravity, showing that
the contribution of the heavy modes regularizes the total energy,
making it finite and positive, despite the non-Newtonian behavior of
the static gravitational potential. This fact has profound
consequences on the interpretations of these solutions. In fact, one
may accept a slow fall-off of the new term, with $-1<\g<2$, and thus
allow for a non-Newtonian gravitational force, still with a finite
total energy.

\SEC{Gravitational Energy in GR.} Let us briefly review the notion of
quasilocal energy in gravity theories.  In the past, many proposal of
quasilocal energies have been put forward, in fact it has now been
recognized that there is an infinite number of them with reasonable
physical properties.  According to the Noether charge method, in a
general covariant theory, for each diff one can construct a conserved
charge that is expressed as an integral of the relative
generators. For this to construction to work, it is necessary that all
fields that are varying under a diff be dynamical, i.e.\ that there is
no external, fixed background.  The energy will then be given by the
value of the Hamiltonian, the generator of the translations in time.
Following the analysis of Chen and Nester~\cite{nesterm}, each
conserved charge $Q(\xi)$ is made of a bulk term (integrated over a
$t=$const. hypersurface $\Sigma_t$) and a surface term (integrated
over the boundary surface $\de \Sigma_t$)
\be
Q(\xi) = \int_{\Sigma_t} d^3x \,  {\cal H}(\xi) + \int_{\de \Sigma_t}
d^2x \,  {\cal B}(\xi)  \, .
\ee
For diffeomorphism invariant theories the bulk density is vanishing on
shell, thus the only contribution to $Q(\xi)$ comes from the boundary
$\de \Sigma_t$.

The surface term has to satisfy few requirements
\begin{itemize}
\item The conserved charge has to be functionally differentiable: the
  variation $\delta Q$ is well defined if all surface terms vanish.
  That is the surface terms from the variations in the bulk should
  exactly cancel out the variations in the boundary term, when
  suitable boundary conditions are used.
\item The boundary term must be covariant.
\end{itemize}
The last requirement is fulfilled by Dirichlet and Neumann boundary
conditions and also requires a reference background, often taken to be
Minkowski space.  Actually, as shown by Kijowski ~\cite{kt,kijowski},
the choice of a reference background is crucial for a proper
definition of the symplectic space of GR and it is equivalent to
fixing the boundary conditions or the asymptotics for the canonical
variables.  For a scalar field, both Dirichlet and Neumann boundary
condition correspond to adiabatic insulation of the system; indeed, no
energy flux across the spatial boundary is allowed.  Neumann and
Dirichlet are however the only boundary conditions that allow a
covariant definition of the symplectic structure of gravity; in
addition, in the Dirichlet case, the corresponding energy is positive
definite~\cite{nester1}, and gives at null infinity the Bondi
mass~\cite{nest-inf}.

Using Dirichlet boundary conditions for the metric on  $\de \Sigma_t =
{\cal S}_t$, the quasilocal energy is given by  
\be
\begin{split}
&E = \frac{1}{32 \pi G} \int_{S_t} d^2z \, 
\epsilon_{\rho \sigma \mu\nu}  \\ 
&\left(\xi^\tau \Pi^{\beta \lambda}
  \Delta \Gamma^\alpha_{\beta \gamma} \, 
\delta^{\mu \nu \gamma}_{\alpha \lambda \tau} + \bar \nabla_\beta \xi^\alpha \Delta
  \Pi^{\beta \lambda} \, \delta^{\mu \nu}_{\alpha \lambda} \right) 
 \frac{\de x^\rho}{dz^1}   \frac{\de x^\sigma}{dz^2} \, ,
\end{split}
\label{nqe}
\ee 
where $z^a$ are local coordinates on ${\cal S}_t$, $\Pi^{\mu \nu} =
\sqrt{g}\, g^{\mu \nu}$ and $\Gamma^\alpha_{\mu \nu}$ is the
Levi-Civita connection in a coordinate basis. The difference between a
quantity $f$ and its background value $\bar f$ is denoted by $\Delta
f$.  Thus the quasilocal energy will depend on the reference metric
$\bar g$ and the timelike vector associated with time flow. As matter
of fact, the energy (\ref{nqe}) coincides with the one found by Katz
at al.~\cite{Katz} using a completely different approach.  For our
purposes it is convenient to consider the following metric
\be
ds^2 = -{\cal J}(r) \,  d  t^2 +{\cal K} (r)\,  dr^2 + 2{\cal D}(r) \,
dt dr  + \bar B \, r^2
\, d \Omega^2 . 
\ee
We take
as reference the flat background
\be
d \bar s^2 =- \bar{\cal J} \, d  t^2 + \bar B \left( dr^2  + 
\, r^2 d \Omega^2 \right) \, , 
\ee
with $\bar{\cal J}$ and $\bar B$   constants. Thus,
if  the time flow vector is $\xi= \xi^0(r) \, \frac{\de}{\de t}$, the energy is given by
\be
\! \!
E = \frac{\bar B\,  r}{2\,  G \, }\!\! \left[\frac{ \xi^0}{  {\cal F}^{1/2}}\!
 \left(  \frac{\cal F}{\bar  B}- {\cal J}\right) 
- r \, {\xi^0}' \!\!  \left(  \frac{ \bar{\cal J}}{ \bar{\cal F}^{1/2}} -  \frac{{\cal J}}{ {\cal F}^{1/2}}  \right)\right] ,
\label{mast}
\ee
where we have set ${\cal F} = {\cal J K } + {\cal D}^2$ and $\bar{\cal
  F}$ is the corresponding quantity for the background metric.  For
the standard Schwarzschild metric ${\cal J}= 1/{\cal K}=1-2\,GM/r$,
${\cal D}=0$ and $\bar{\cal J}= \bar B=1$. When we set $\xi^0=1$ (the
time flow coincides with the Killing vector) the energy coincides with
the mass of the black hole $M$.

\SEC{Massive Gravity and Exact BH Solutions.}  An elegant way to
provide the needed new DoF for the massive phase of gravity, without
breaking diff invariance, is to work with an extra tensor $\tilde
g_{\mu \nu}$, with a kinetic term similar to GR, this bring us into
the realm of bigravity theories whose study started in the '60 (see
\cite{DAM1} for early references).  Once we couple nonderivatively the
extra spin two to the standard metric field we can explore both
Lorentz invariant and Lorentz breaking phases of massive gravity,
working with consistent and dynamically determined backgrounds.  The
action for the modified gravity model we will consider is \be
\begin{split}
S=
\int d^4 x  \,\sqrt{g} 
M_{pl}^2 \, {\cal R}
  + 
{} \left[\sqrt{\tilde g} \, \tilde M_{pl}^2  \, \tilde {\cal R}  -
4  (g \tilde g)^{1/4}  V \right] ,
\end{split}
\label{act} 
\ee
where the interaction potential $V$ is a scalar function of the tensor
$X^\mu_\nu = {g}^{\mu \alpha} {\tilde g}_{\alpha \nu}$.  Clearly the
action is invariant under diffs.

For a large family of potentials the equations of motion admit the
following spherically symmetric solutions~\cite{Berezhiani:2008nr}
(where $M_{pl}^{-2} = 16 \pi G$ and $\kappa =\tilde M_{pl}^2
/M_{pl}^{2} $):
\begin{itemize}
\item For the metric $g_{\mu\nu}$
\bea
\!\!\!ds^2&=&-J(r)\, dt^2 + J^{-1}(r)\, dr^2 +r^2\, d\Omega^2\,,\nonumber
\\
J(r) &=& 1+ 2\,G \left(-\frac{M}{r}+ S\,r^\g\right) 
\label{sol1}
\eea
\item For the metric $\tilde g_{\mu\nu}$
\bea
\!\!\!d \tilde s^2&=& -\tilde J(r)\, dt^2  +
\tilde K(r)\, dr^2  + 2 \tilde D(r)\,dt\,dr +\bar B\,
r^2 d \Omega^2, \nonumber
\\
\tilde J (r) &=& c^2\o^2\left[1+2  \, \frac{G}{\kappa\,\o^2} \left(-\frac{ c \,\tilde M}{ r}-\frac{ S}{c }\,r^\g\right)\right],\nonumber\\[1ex]
\bar B&=& \o^2 \,,\qquad \tilde K(r)=(c^2\o^4+\tilde D^2(r))/\tilde J(r)\,,
\label{sol2}
\eea
\end{itemize}
\vspace*{-2ex}%
where $c$, $\o$ and $\g$ are coefficients depending on the potential
$V$.  In these solutions, besides the standard term $M/r$, there is a
power-like modification with a new integration constant $S$ which sets
its size.\footnote{When $\gamma = -1$, the modification is
  logarithmic, $ r^\gamma \to \log r/r$.
} When $\gamma < 2$, both metrics describe asymptotically flat spaces
(${\cal R}_{\mu\nu\rho\sigma}\sim r^{\g-2 }$), namely Minkowski space
($\eta_{\mu\nu}$) for $g_{\mu \nu}$ and the metric $\tilde \eta_{\mu
  \nu} =\o^2 \, \text{Diag}( -c^2 , \, 1 , \, 1, \, 1 )$ for $\tilde
g_{\mu \nu}$.  Notice that when $c \neq 1$, asymptotically, Lorentz
symmetry is broken down to rotations~\cite{usLB}.  The solution for
$\tilde g_{\mu\nu}$ is quite similar to the one for $g_{\mu\nu}$,
however an off-diagonal $t-r$ contribution $\tilde D$ is present.  Its
explicit form is not essential, but it is important to recall that at
large distances its behavior is $\tilde D\sim 1/\sqrt{r}$ (for $\gamma
< -1$).  This is slower than the typical newtonian weak-field behavior
and is analogous to the Schwarzschild solution written in Painlev\'e
coordinates. Contrary to that case, here this term cannot be gauged
away; in fact, by the change of coordinates $dT= dt - \tilde D/\tilde
J \, dr$, one can bring $\tilde g_{\mu\nu}$ to diagonal form, but then
$g_{\mu\nu}$ is non diagonal. Incidentally, the asymptotic behavior
of $D$ implies that the ADM energy found in the standard Hamiltonian
formulation of GR~\cite{ADM,RT} is not well defined~\cite{oMurchada}.

A similar solution for $g_{\mu \nu}$ exists also in the St\"uckelberg
effective approach~\cite{Comelli:2010bj,Bebronne:2009mz}; this time
$\tilde g_{\mu \nu}$ is replaced by a set of four scalar fields.  In
the decoupling limit of $\tilde g_{\m\n}$ ($\kappa \to \infty$) the
fluctuations of the auxiliary metric are decoupled, and $\tilde g_{\mu
  \nu}$ is practically frozen to its background value $\tilde
\eta_{\mu \nu}$, modulo gauge. From~(\ref{sol1}), (\ref{sol2}) one can
see that in this limit, the new term proportional to $S$ survives in
$g_{\m\n}$, while $\tilde g_{\m\n}$ becomes
flat~\cite{Berezhiani:2008nr}. A preliminary study of the
thermodynamics of these black hole solutions can be found
in~\cite{BHT}.

\SEC{Total Energy.} The computation of the total energy of the system
described by (\ref{act}) follows the same steps as in GR. The crucial
point is that the interaction part in the action does not provide new
boundary terms in the total Hamiltonian (due to the fact that the two
sectors interact without derivative terms).  As a result, the diffs
Noether charge can be written, on-shell, as the sum
\be
Q_{\text{tot}}(\xi) = \int_{{\cal S}_t}
d^2x \,  \left[ {\cal B}(\xi) + \tilde {\cal B}(\xi) \right] \, ,
\ee
where $ {\cal B}(\xi)$ and $\tilde {\cal B}(\xi)$ are the single
contributions from the kinetic action of $g$ and $\tilde g$
respectively.  The energy of the solutions (\ref{sol1}) and
(\ref{sol2}) can be worked out by using the general formula
(\ref{mast}) applied to $g$ and $\tilde g$.  For $g$ we take as
reference space the Minkowski space $\eta_{\mu\nu}$, while for $\tilde
g$ the natural reference space is its diagonal vacuum metric $\tilde
\eta$.  Taking as usual $\xi = \frac{\de }{\de t}$, the contributions
to the total energy coming from $g$ and $\tilde g$ are
\bea
\label{E1}
E&=&\frac{r}{2  \, G  } \left(1- J \right) ={M}-S\,r^{\g+1},\\
\tilde E &=& \frac{r \kappa}
{2 \, c \, G}\left(c^2 \omega ^2-  \tilde J\right) =\tilde M\,c^2+S\,r^{\g+1}\,.
\label{E2}
\eea
The term $S\, r^{\gamma+1}$ in both $E$ and $\tilde E$ would lead to a
divergent energy in the limit $r \to \infty$ when $-1 <\gamma < 2$, as
in the newtonian case.  Amazingly instead, the true total energy is
finite: the power-like terms proportional to $S$ cancel exactly,
leading to
\be
E_{tot}= E+\tilde E = {M} + \tilde M \, c^2\,.
\label{sum}
\ee
The system has thus finite energy, given just by $M$ and $\tilde M$
from the Newtonian terms.

Let us observe at this point that in the effective-theory approach the
picture is drastically different. As we mentioned, taking first the
limit $\kappa \to \infty$ in the solution~(\ref{sol1}) (\ref{sol2}),
the auxiliary metric gets frozen to its flat reference background
$\tilde \eta$, while the physical metric $g$ is unmodified. Therefore
one would attribute zero energy to the auxiliary metric and now the
total gravitational energy appears to be divergent when $\gamma >-1$.
In the decoupling limit thereore, an infinite amount of energy is
neglected, spoiling completely the nice cancellation mechanism.  This
also shows that the computation of the energy and the decoupling limit
do not commute.  Clearly, the cancellation can take place only if the
auxiliary metric is dynamic.\footnote{We also checked that, for the
  less physical Neumann boundary conditions~\cite{nesterm}, where
  $\delta \Gamma^\alpha_{\mu\nu}=0$ at the boundary, the corresponding
  energy $E_{tot}^{Neu}$ is in general not finite anymore. When
  $\gamma < -1$ one still recovers the finite total energy (\ref{sum})
  at large $r$, however $E_{tot}^{Neu}$ turns out to be divergent if
  $\gamma > -1$.  In particular $E_{tot}^{Neu} \sim r ^{\gamma + 1}$
  for $-1 < \gamma < 0$ and $E_{tot}^{Neu} \sim r ^{2 \gamma + 1}$
  when $0 < \gamma < 2$.}

\medskip

It is also interesting to study what happens when the total energy is
expressed in terms of the body internal structure. For the sake of
simplicity we considered a body of radius $R_\odot $, made of constant
density fluid~\cite{Berezhiani:2008nr}, with total mass $ m =
\frac{4}{3}\pi R_{\odot}^3 \, \rho$.  The integration constants $M, \,
\tilde M$ and $S$ can be determined by matching equations
(\ref{sol1}), (\ref{sol2}), representing the external part of the
solution (valid for $r>R_{\odot}$ where $R_{\odot}$ is the body's
radius) with the interior one.  The result of the matching is
\be
{M}=m+\Delta M,\qquad
\tilde M=
-\,\frac{\Delta M}{c^2} 
\label{ren}
\ee
and
\bea
S= b  \,  \mu_{grav}^2 m \, R_{\odot}^{1-\g},\qquad
\Delta M = \mu_{grav}^2\, m \, R_\odot ^{2} \, ,
\eea
with $b$ a suitable constant not relevant for the present discussion
(see~\cite{Berezhiani:2008nr}). The mass parameter $\mu_{grav}^2$ is
determined by the potential $V$: generically, it has indefinite sign
and is of the same order of the graviton mass. The nonzero value of
$S$ means that the modification of the potential is turned on by the
matter source $m$, and the appearance of $\Delta M$ in (\ref{ren})
shows that the body mass $m$ in the Newtonian term is renormalized,
leading to screening/antiscreening of mass. This happens because of
the gravitational interaction between $g$ and $\tilde g$.

Nevertheless, evaluating eq.~(\ref{sum}), we get that the total
energy coincides with the ``bare'' mass $m$,
\be 
E_{tot}={M}+\tilde M \, c^2=m\,.
\label{tot}
\ee
This result is striking, since the total energy is independent not
only from $S$ but also form the mass renormalization $\Delta M$,
suggesting that some sort of Gauss Law is still valid for the total
energy.

\medskip

\pagebreak[3]

\SEC{Conclusions.} We studied the total gravitational energy of the
exact spherically symmetric solutions, in a class of massive gravity
theories where the metric field couples in a nonderivative way to an
additional tensor field.  In such solutions the gravitational
potential features an interesting modification $\Phi \sim {M}/r + S
r^{\g}$, which describes an asymptotically flat spacetime for $\gamma
< 2$.  Clearly, when $-1 < \gamma <2$, the asymptotics is peculiar and
in conflict with the standard behavior $\Phi\sim 1/r$ for $r\to
\infty$, expected in a system with finite energy. Nevertheless, we
have shown that the new field responsible for the modification also
contributes to the {\rm total} energy that turns out to be finite and
positive.  The point is that the seemingly infinite amount of energy
due to the $r^\gamma$ term is cancelled by a similar contribution from
the backreaction of the additional field.

The new field is coupled to gravity only, and while it will affect the
propagation of gravitational waves~\cite{usLB}, it cannot be probed by
standard matter experiments. The standard observers which couple to
metric $g_{\mu \nu}$ will not experience any violation of the weak
equivalence principle, but they will observe the peculiar behavior of
the gravitational potential $\Phi$ at large distances.  Just by
looking at the asymptotics of $\Phi$, they would be baffled by its
behavior and led to conclude that the ``total'' gravitational energy
of the system is infinite, while in reality it is not.  The total
energy is not only finite but also does not depend on the constant $S$
appearing in the solution. Though the BH-like solution has hairs (the
values of $S$) they have no effect on the total energy.

When the solution describes the exterior part of a self-gravitating
body, e.g.\ a realistic star, the integration constants $M$, $\tilde
M$ and $S$ can be computed in terms of the body's density and
radius. While both $M$ and $\tilde M$ show in general (anti)screening
of mass, and together with $S$ are in general complicated functions of
the body shape (size), the cancellation mechanism is again operative,
and as a result, the total energy content of a stellar object is just
its ``bare'' mass.

The picture that emerges is that in modified gravity theories the
gravitational potential can greatly differ from our naive
expectations. The potential can have a softer than expected fall-off,
without clashing with the basic principle that any finite physical
system must have finite energy.  This we think is remarkable and can
change our view on gravity especially if the cancellation mechanism is
a general feature of modified gravity.

It will also be very interesting to explore the cosmological
implications of the scenario described in this work, and we leave it
for a future investigation. Clearly, an important open question is the
stability of our solutions, in particular when the term $r^\g$
dominates at large distances ($-1<\g<2$).  The answer is not an easy
one, entailing the nontrivial study of perturbations around the
solution (\ref{sol1})-(\ref{sol2}), and clearly deserves further
study.
 
\bigskip

\SEC{Acknowledgments.}  We thank P. Menotti for useful discussions.
D.C. and L.P. thank ICTP for hospitality during the completion of the
work.

\end{document}